\documentclass[twocolumn,pre]{revtex4}

\usepackage{natbib}
\usepackage{multirow}
\usepackage{amsmath}
\usepackage{graphicx}

\newcommand{\hc}{H_C}
\newcommand{\hod}{H_{OD}}
\newcommand{\hld}{H_{LD}}

\newcommand{\rk}{r_{\mathrm{deg}}}
\newcommand{\Ct}{C_{\textrm{target}}}
\newcommand{\Cg}{\widetilde{C}}

\newcommand{\gi}{\gamma_i}

\newcommand{\defn}{\textbf}

\newcommand{\eref}[1]{(\ref{#1})}
\newcommand{\av}[1]{\langle#1\rangle}

\newcommand{\e}{\mathrm{e}}
\renewcommand{\d}{\mathrm{d}}

\newcommand{\set}[1]{\{#1\}}

\newcommand{\half}{{1\over2}}
\newcommand{\etal}{\textit{et al.}}

\newcommand{\cN}{\mathcal{N}}

\newcommand{\ex}[1]{\e^{#1}}

\newlength{\figurewidth}
\ifdim\columnwidth<20cm
\else
\fi
\begin{document}
\title{Linear and Optimization Hamiltonians in Clustered Exponential Random Graph Modeling}
\author{Juyong Park}
\affiliation{Department of Physics, Kyunghee University, Seoul, Korea}
\author{Soon-Hyung Yook}
\affiliation{Department of Physics, Kyunghee University, Seoul, Korea}

\begin{abstract}
Exponential random graph theory is the complex network analog of the canonical ensemble theory from statistical physics. While it has been particularly successful in modeling networks with specified degree distributions, a na\"ive model of a clustered network using a graph Hamiltonian linear in the number of triangles has been shown to undergo an abrupt transition into an unrealistic phase of extreme clustering via triangle condensation. Here we study a non-linear graph Hamiltonian that explicitly forbids such a condensation and show numerically that it generates an equilibrium phase with specified intermediate clustering.
\end{abstract}
\maketitle

\section{Introduction}
\label{introduction}
The study of complex systems found in various disciplines including engineering, biology, sociology that can be represented as networked systems composed of nodes and edges have garnered much interest from statistical physicists in recent years. Building upon a rich and long tradition of studies on many-body systems, they have successfully adapted the analytical and computation tools to understanding networks.~\cite{Dorogovtsev:2003,Albert:2002rmp,Newman:2008}

\begin{figure}[t]
\includegraphics[width=80mm]{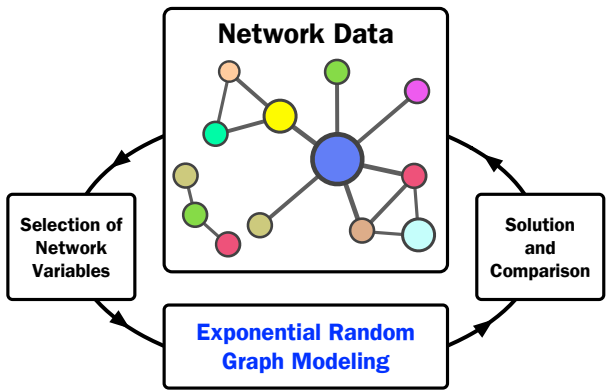}
\caption{The schematics of the Exponential Random Graph modeling of network data. From the network data of interest (top) one selects network variables such as the node degrees $\set{k_i}$ (left) from which one then forms a Hamiltonian $H(\set{k_i})$, whose solutions and predictions are compared with the network data. Significant disagreements may necessitate a new selection of variables or reformulation of the Hamiltonian.}
\label{00_procedure}
\end{figure}

A network modeling methodology that shows a striking resemblance to the canonical ensemble theory from statistical physics is the Exponential Random Graph (ERG) theory, originally developed in statistics and currently the most actively studied in the Social Network Analysis (SNA) circles~\cite{Frank:1986,Park:2004sm,Robins:2007}. Given that the potential readership of this paper will be composed of statistical physicist, the premise of ERG is perhaps most simply explained using the language of statistical physics. Here, as in the canonical ensemble theory, one considers an ensemble $\Gamma$ of graph configurations (microstates) $G$ whose probabilities in $\Gamma$ are given by $P(G)=\sum_{G\in\Gamma}\ex{-H(G)}/Z$ where $H(G)$ is the \emph{graph Hamiltonian}, a function of network characteristics of $G$, and $Z=\sum_{G\in\Gamma}\ex{-H(G)}$ is the partition function. Both in social network analysis and statistical physics, the Hamiltonian $H(G)$ is typically set up to be a linear function of \emph{network variables} or \emph{network statistics} such as the number of edges $m(G)$ in the graph. When the network is simple and unweighted (i.e. the number of edge between two nodes is either $0$ or $1$) it is straightforward to show that $H(G)=\theta m(G)$ generates the so-called Erd\"os-R\'enyi random graph in which two nodes are connected with probability $p=1/(1+\ex{\theta})$~\cite{Park:2004sm}. The expected number of edges $\overline{m}$ in a network of $n$ nodes is in this case, therefore, given as
\begin{align}
\overline{m} &= {n\choose 2}p = \frac{n(n-1)}{2}\frac{1}{1+\ex{\theta}},
\label{eredges}
\end{align}
controlled by the conjugate variable $\theta$. In one then equates $\overline{m}$ from Eq.~\eref{eredges} with the actual number of edges $m$ in the network data under study, this serves as the \emph{null model} of the network under study which the number of edges as the only observable. Note again that only the number of edges $m$ is an explicit variable in constructing the network ensemble~\footnote{Typically we consider the number of nodes $n$ as given.}; whether the model is sufficient (i.e., is a good approximation of network data) is to be judged on the given model's ability to reproduce other (not used as an input to the model) network characteristics such as the degree distribution, cluster size distribution, degree-degree correlation,~\emph{et cetera}. A significant disagreement between the expected characteristic of a model and the data may indicate that the choice Hamiltonian needs to be reformulated; for instance, the ubiquity of scale-free (power-law) networks where the degree distribution is fat-tailed renders the simplest Erd\"os-R\'enyi network model (which has a Poissonian degree distribution) inadequate, necessitating the introduction of alternative forms of the graph Hamiltonian. One possibility is to incorporate explicitly the node degrees e$\set{k_i}$ ($i\in\cN={1,\cdots,n}$ is the node indices) themselves to form the so-called \defn{Linear Degree Hamiltonian} $\hld(G)$:
\begin{align}
H_{LD}(G)=\theta_1k_1(G)+\cdots+\theta_nk_n(G),
\label{ldham}
\end{align}
where $\set{\theta_i}$ are the conjugate variables that now control the expected degrees $\set{\overline{k_i}}$ in a manner similar to what $\theta$ did to $\overline{m}$ in Eq.~\eref{eredges}~\footnote{Note the absence of the temperature $\beta$ in Eqs.~\eref{eredges}~and~\eref{ldham}. Here, one may consider $\beta$ as having been absorbed into $\set{\theta}$.}. On a historical note, the study of $\hld$ was prompted by the hypothesis that heavily skewed degree distributions such as the power law may cause the observed negative correlation between degrees of connected nodes, while the Erd\"os-R\'enyi network produces no such correlation in the thermodynamic limit ($n\to\infty$)~\cite{Park:2004sm}. On the other hand, it was shown analytically that power-law networks generated via Eq.~\eref{ldham} exhibited negative degree correlation, proving the hypothesis, and thus that the skewed degree distribution was indeed responsible for the negative degree-degree correlation. This is, in fact, a typical example of the ERG modeling (also of general statistical modeling procedure) procedure   -- identifying ``important'' features of the observed system and testing its sufficiency via comparing the model's predictions and real data (i.e. the ``goodness of fit'' of the model in statistical sense. See Ref.~\cite{Hunter:2008}) and, when a closer agreement is desired, refining the hypothesis and repeating the procedure. This process is presented schematically in Fig.~\ref{00_procedure}.

Not surprisingly, the development of ERG as a network modeling framework closely follows the study of graph Hamiltonians of increasing complexity. ERG models of historical import include, in addition to the simplest $H(G)=\theta m(G)$, the Holland and Leinhardt model of reciprocity, the Strauss model of clustering, the 2-star model, and the generalized $k$-star models~\cite{Park:2004sm}. We refer interested readers to introductory articles and significant recent work from the SNA community for more detail~\cite{Anderson:1999,Robins:2007,Hunter:2007,Hunter:2008}.

To a statistical physicist, the benefits of such a formalism is obvious: one can utilize appropriate computational (such as the Metropolis-Hastings algorithm) and analytical (such as Feynman-diagrammatic method) tools to study the properties of the model~\cite{Metropolis:1953,Park:2004sm,Park:2010}. It should also be noted that the Hamiltonian need not be linear at all. For instance, when one wishes to construct an exponential random graph model of a network with a specified degree sequence, $H(G)$ only needs to be a function of the node degrees $\set{k_i(G)}$ in $G$, i.e. $H(G)=H[\set{k_i(G)}]$ where $i\in\cN=\set{1,\ldots,n}$ is the node index. This is sufficient to guarantee that two configurations $G$, $G'$ with an identical degree sequence have the same probability in the ensemble, and the aforementioned $\hld$ is one possibility. Thus there is much freedom in choosing the form of the $H(G)$, meaning there exists ample avenues for exploration of various possible forms of Hamiltonians as one sees fit, not limited to linear forms. In fact, linear forms such as $\hld$ of Eq.~\eref{ldham} are often not robust in the presence of a perturbation, in the sense that when a composite Hamiltonian $H=\hld+H'$ is used the equilibrium degree distribution may differ significantly from the one specified from $\hld$, defeating the modeler's intention to generate a desired degree distribution using $\hld$. The purpose of this paper is to review the clustering perturbation and compare the characteristics of linear and nonlinear Hamiltonians under it. For simplicity, we here consider only unweighted and undirected graphs.

\section{Degree Hamiltonians}
Here we briefly review $H_{LD}(G)=\sum_i\theta_ik_i$, Eq.~\eref{ldham}, specifically when the network is \emph{sparse} ($k_i\sim O(1)\ll \sqrt{n}$). In such a case it is well known that the probability $p_{ij}$ that nodes $i$ and $j$ are connected is $\ex{-\theta_i}\ex{-\theta_j}$, leading to the average degree $\av{k_i}$ of node $i$~\cite{Park:2004sm}
\begin{align}
\av{k_i}
	&= \sum_{j\ne i}p_{ij} = \ex{-\theta_i}\sum_{j\ne i}\ex{-\theta_j}\nonumber \\
	&=(n-1)\ex{-\theta_i}\int_{-\infty}^{\infty}\ex{-\theta}\rho(\theta)\d\theta\equiv A(n-1)\ex{-\theta_i},
\label{avk}
\end{align}
where the latter integral form is valid for a large network ($n\gg1$), $\rho(\theta)$ is the distribution density of $\theta$, and $A\equiv\int_{-\infty}^{\infty}\ex{-\theta}\rho(\theta)\d\theta$ is thus a constant. Setting $\av{k_i}=q_i$, the specified (desired) degree of node $i$ and inverting Eq.~\eref{avk}, we obtain $\theta_i = -\ln\bigl(q_i/A(n-1)\bigr)$. $\hld$ then becomes
\begin{align}
\hld(G)	&= \sum_{i\in\cN}\theta_ik_i(G) \nonumber \\
	&= -\sum_{i\in\cN}k_i(G)\ln q_i+\ln \bigl(A(n-1)\bigr)\sum_{i\in\cN}k_i(G) \nonumber \\
	&= -\sum_{i\in\cN}k_i(G)\ln q_i+2M(G)\ln \bigl(A(n-1)\bigr),
\label{linham1}
\end{align}
where $M(G)=\half\sum_ik_i(G)$ is the number of edges in $G$. One can also show that the ensemble generated via $\hld$ is equivalent to the \emph{configuration model}, a popular and useful framework for studying graphs with arbitrary degree distributions~\cite{Newman:2001arb}.

Now, if we restrict the ensemble $\Gamma=\set{G}$ to contain only network configurations $G$ with a fixed number of edges $M(G)=M_0=\half\sum_iq_i$ (corresponding to the canonical ensemble of particles), the second term $2M(G)\ln A(n-1))$ becomes a constant. Therefore, we can safely ignore it and use an even simpler form
\begin{align}
\hld(G) = -\sum_{i\in\cN}k_i(G)\ln q_i.
\label{linham2}
\end{align}
This is particularly useful in edge-conserving Monte Carlo simulations, where the Metropolis-Hastings algorithm would consist of relocating the edge between a randomly selected connected node pair to between a randomly selected unconnected pair with probability $1$ if it results in a lower energy, and with probability $\ex{-\Delta H(G)}<1$ when it results in a higher energy.

\begin{figure}[t]
\includegraphics[width=80mm]{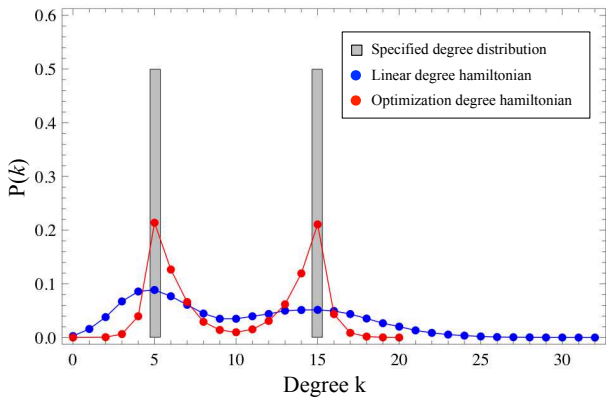}
\caption{The degree distributions from exponential random graph simulations. For simplicity we set the specified degree distribution to be $P(q=5)=\half$ and $P(q=15)=\half$ (shown in gray). The linear degree Hamiltonian $\hld=-\sum_{i\in\cN}k_i\ln q_i$ generates a smooth distribution over a wide range of degrees with Poissonian-like peaks at $k=5$ and $k=15$ (blue). The degree distribution from the optimization degree Hamiltonian $\hod=\sum_{i\in\cN}|k_i-q_i|$, by contrast, is noticeably closer to the specified, with sharper peaks of roughly equal heights at $k=5$ and $k=10$.}
\label{01_degree}
\end{figure}

It is important to note that it is the ensemble average $\av{k_i}$ of a node that is to be matched with its prescribed degree $q_i$, and there is no guarantee that $k_i=q_i$ strictly, even at equilibrium: In fact, $P(k_i|q_i)$, the probability that a node with a prescribed degree $q_i$ has degree $k_i$ at equilibrium, is
\begin{align}
P(k_i|q_i) = \sum_{\set{\cN_k}}\biggl[\prod_{j\in \cN_k}\ex{-(\theta_j+\theta_i)}\prod_{l\in\cN_k'}\bigl(1-\ex{-(\theta_l+\theta_i)}\bigr)\biggr],
\label{pkq}
\end{align}
where $\theta_i=-\ln q_i/A(n-1)$, and $\set{\cN_k}$ is the set of all possible combinations of $k$ nodes from $\cN$ excluding $i$. From this, the total degree distribution $P(k)$ in equilibrium is given as
\begin{align}
P(k) = \sum_{\set{q}}P(k|q)P(q),
\label{linPk}
\end{align}
where $P(q)$ is the prescribed degree distribution. It is unlikely that $P(k=q|q)\equiv1$ in Eq.~\eref{pkq}, and thus we can not expect $P(k)\equiv P(q)$.  To find the general characteristics of $P(k)$ from Eq.~\eref{linPk} in comparison with $P(q)$, we performed a Monte Carlo simulation (using the Metropolis-Hastings method described above) of $\hld$ for a network of $n=500$ and $P(q=5)=P(q=15)=\half$ for illustrative purposes, whose results are shown in Fig.~\ref{01_degree}. In the figure, the prescribed $P(q)$ is shown in gray, and the equilibrium $P(k)$ is shown in blue. While $P(k)$ does exhibit peaks at $k=5$ and $k=15$, it also shows a fairly wide distribution (although small in comparison with $n$), and the fluctuation is visibly larger at $k=15$ resulting in a lower peak.  If the specified degree $q$ had been the same for all nodes (i.e. a $q$-regular graph) it is well known that $\hld$ creates an Erd\"os-R\'enyi graph with a Poissonian degree distribution, locally not unlike the peaks in Fig.~\ref{01_degree}~\cite{Park:2004sm}. Thus we call the peaks we see in Fig.~\ref{01_degree} Poisson-like.

\begin{figure*}[t]
\includegraphics[width=140mm]{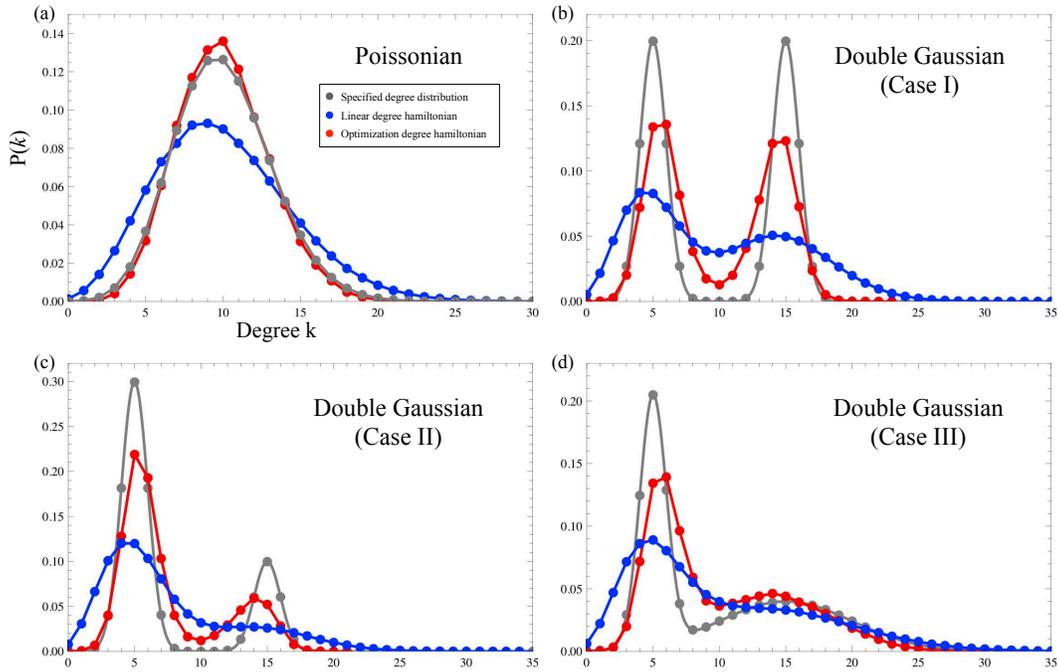}
\caption{The degree distributions generated via $\hld$ and $\hod$ for heterogeneous specified degree distribution. (a) With $P(q)$ a Poissonian (with $\av{q}=10$). In (b)--(d) $P(q)$ is a double Gaussian with peaks at $q=5$ and $q=10$ with varying relative heights ($\alpha\in[0,1]$ for the peak at $q=5$, and $1-\alpha$ for the peak at $q=10$) and variances $\sigma_1$, $\sigma_2$ of the peaks. (b) $(\alpha,\sigma_1,\sigma_2)=(0.5,1.0,1.0)$. This is the most similar to Fig.~\ref{01_degree}. (c) $(\alpha,\sigma_1,\sigma_2)=(0.75,1.0,1.0)$ and (d) $(\alpha,\sigma_1,\sigma_2)=(0.5,1.0,5.0)$. Here, $\hod$ again consistently reproduce $P(q)$ more faithfully.}
\label{01A_smoothdegree}
\end{figure*}

The well-documented success of the configuration model implies that the fluctuations we see in $P(k)$ may not be problematic in general, though in certain circumstances (we see later such as a case) a more faithful reproduction of the specified degree distribution may be desired. This means that a graph Hamiltonian is needed that imposes a larger penalty when $k_i$ deviates from $q_i$ than $\hld$ does.  It is unclear how $\hld$ can be modified while retaining the linear form. Instead, we introduce a nonlinear Hamiltonian
\begin{align}
H_{OD}(G) = \sum_{i\in\cN}\beta_d|k_i(G)-q_i|
\label{optham}
\end{align}
which we call the \defn{Optimization Degree Hamiltonian}, being reminiscent of Hamiltonians used in certain optimization problems such as number partitioning~\cite{Mertens:1998sd}~\footnote{Note that the so-called ``curved'' Exponential Random Graph Model is similar to our formalism in that the graph Hamiltonians are nonlinear functions of network variables~\cite{Hunter:2008}.}. The $P(k)$ that results from $H_{OD}$ with $\beta=1$ for simplicity (the penalty can be controlled via $\beta_d$ when necessary) is shown in Fig.~\ref{01_degree} in red, which is indeed a more faithful reproduction of $P(q)$ in comparison with $\hld$, showing sharper peaks at $k=5$ and $k=15$ of equal heights similar to $P(q)$. The broadening of the peaks around the specified degrees from the $\hld$ in comparison to $\hod$ in Fig.~\ref{01_degree} is persistent in cases of more heterogeneous (thus less artificial) specified $P(q)$ as seen in Fig.~\ref{01A_smoothdegree} where we compare $\hld$ and $\hod$ for a Poissonian $P(q)$ and a double Gaussian
\begin{align}
P(q)=\alpha \Phi(q;\mu_1,\sigma_1)+(1-\alpha)\Phi(q;\mu_2,\sigma_2)
\label{Pqdoublegauss}
\end{align}
where $\Phi(q,\mu,\sigma)$ is a Gaussian of mean $\mu$ and variance $\sigma^2$, and $\alpha\in[0,1]$ sets the relative weights between the two Gaussian peaks. For the Poissonian case we set $\av{q}=10$ (Fig.~\ref{01A_smoothdegree}~(a)), and for the double Gaussian we try three cases of varying weights and variances (Fig.~\ref{01A_smoothdegree}~(b)-(d)). The behavior of $\hld$ and $\hod$ are consistent with what we see from Fig.~\ref{01_degree}: in terms of the goodness of fit to $P(q)$ (including the relative heights at the peaks) $\hod$ is superior to $\hld$~\footnote{For the purposes of this paper we are showing some numerical examples. For more systematic studies one could investigate various moments of the degree distributions from the two Hamiltonians, or a difference measure between two distributions $P$ and $Q$ such as $D(P,Q)\equiv\sum_k|P(k)-Q(k)|$.}.

\section{Robustness of Degree Reproduction under perturbation: targeted clustering}
Besides degree distribution, a network characteristic that has been widely studied is clustering. Intuitively, a clustered network contains significantly more triadic closures (triangles) than expected in a random graph with the same number of edges. (A common definition of the strength of clustering of a network is given using the so-called \emph{clustering coefficient}, which we present later). 

In exponential random graph literature, studies have been made on graph Hamiltonians that incorporate the number of triangles $T$ linearly, the simplest case being the Strauss Model with $H_S(G)=\theta M+\tau T$~\cite{Strauss:1986}. The motivation for $H_S$ is that by controlling $\theta$ and $\tau$, one could hopefully generate a network with any desired value of $M$ and $T$, i.e. a smooth, controllable transition between a non-clustered configuration (small $T$) and a clustered one (large $T$). Unfortunately, it has been shown that $H_S$ does not show such a behavior: depending on $\theta$ and $\tau$, the system undergoes a first-order phase transition from a sparse ER-like phase with vanishing clustering and a nearly-fully connected phase~\cite{Strauss:1986,Park:2005}, of which most real networks are neither.  More recently, Foster~\etal~performed an extensive study of the Hamiltonian $H(G)=\tau T$ on an ensemble of networks of fixed degree sequences (and thus a fixed number of edges), and found that as $\tau$ is tuned, $T$ shows a series of jumps consisting of first-order phase transitions~\cite{Foster:2010}, each transition indicating the formation of densely connected local cliques.

This pathology renders the linear Hamiltonian for modeling real clustered networks, where the triangles are distributed over the the network without such extreme ``condensation'' of triangles.  The lack of such an intermediate phase in $H_S$ stems from the fact that the addition of a single edge in an already densely connected part of the network can lead to a disproportionately large increase in $T$ and decrease in $H_S(G)$, resulting in the condensed phase energetically favorable. Therefore, it is understood that a Hamiltonian or, more generally, a mathematical formalism is necessary that explicitly discourages such condensation~\cite{Park:2005,Newman:2009}.

Before we find such Hamiltonian in our context of exponential random graphs, let us first review how clustering in networks is quantified. It is often done via the \emph{clustering coefficient} $C$. In wide use are three versions, one local (node-level) and two global (network-wide). On the individual node level, the \emph{local} clustering coefficient is defined as
\begin{align}
C_i \equiv \frac{t_i}{s(k_i)} = \frac{t_i}{\half k_i(k_i-1)}~~~(k_i\le2),
\end{align}
where $t_i$ is the number of triangles of which the node $i$ is at a corner, and $s(k_i)={k_i\choose 2}$ is the number of pairs of neighbors of node $i$, also called a two-star centered on $i$. $C_i$ is therefore the probability that two neighbors of node $i$ are themselves neighbors. The global measure of clustering is commonly given by two measures. One is the average of $C_i$ which we write $\overline{C}$, defined as $\overline{C}\equiv\av{C_i}=\sum_{i\in\cN}C_i/N$, i.e. the average of the local clustering coefficients. The other, which we call $\Cg$, is defined as
\begin{align}
\Cg \equiv \frac{3T}{\sum_{i\in\cN}s(k_i)}=\frac{3T}{\sum_i\half k_i(k_i-1)},
\end{align}
where $T=\frac{1}{3}\sum_{i\in\cN}t_i$ is again the number of triangles in the network. Therefore this is the probability that a randomly selected two-star is a part of a triangle ($3$ exists in the numerator because one triangle contains three two-stars).  Although $\Cg$ and $\overline{C}$ are not identical, $\Cg=\overline{C}$ when $C_i=C_0$ for all $i$. In terms of these quantities, the aforementioned behavior of the Strauss Hamiltonian $H_S=\theta M+\tau T$ can be summarized as the clustering coefficient (local or global) being either $\Cg(\mathrm{or~}\overline{C})\simeq0$ (sparse ER-like phase) or $\Cg(\mathrm{or~}\overline{C})\simeq1$ (condensed phase) or, in other words, $t_i\simeq0$ or $t_i\simeq s(k_i)$ for all $i$ while in a network of intermediate clustering coefficient $C$, $t_i$ would be $\sim Cs(k_i)$. Taking a cue from the latter and Eq.~\eref{optham}, we propose the following nonlinear targeted clustering Hamiltonian
\begin{align}
\hc = \sum_{i\in\cN}\beta_c|t_i-\gamma_is(k_i)| = \sum_{i\in\cN}\beta_c\bigl|t_i-\gamma_i\half k_i(k_i-1)\bigr|,
\label{pertclus}
\end{align}
where $\gamma_i$ is now the specified (i.e. targeted) clustering coefficient for node $i$. The difference between $\hc$ and the model of Milo~\etal~\cite{Milo:2002}, where $H_{\textrm{Milo}}=|T-T'|$ and $T'$ is the specified number of triangles, is that $H_C$ allows us to control local clustering. Similar to $\hod$ of Eq.~\eref{optham}, $\hc$ explicitly penalizes $t_i$ when it diverges from a prescribed value $\gamma_is(k_i)$. In studying the effectiveness of $\hc$ in reproducing the specified local clustering, we would also like to have the option of controlling the degrees $\set{k_i}$ simultaneously.  We have already discussed two Hamiltonians designed specifically for that purpose, $\hld$ and $\hod$.  In the remainder of this paper, therefore, we study the following two composite Hamiltonians
\begin{align}
	H_1 &= \hld+H_C = \sum_{i\in\cN}\bigl[-k_i\ln q_i+\beta_c|t_i-\gamma_i s(k_i)|\bigr] \\
	\textrm{and} \nonumber \\
	H_2 &= \hod+H_C = \sum_{i\in\cN}\bigl[\beta_d|k_i-q_i|+\beta_c|t_i-\gamma_i s(k_i)|\bigr]
\label{compham}
\end{align}
to find out whether either is capable of generating network ensembles exhibiting both the specified degrees and local clustering. 

\begin{figure}[t]
\includegraphics[width=80mm]{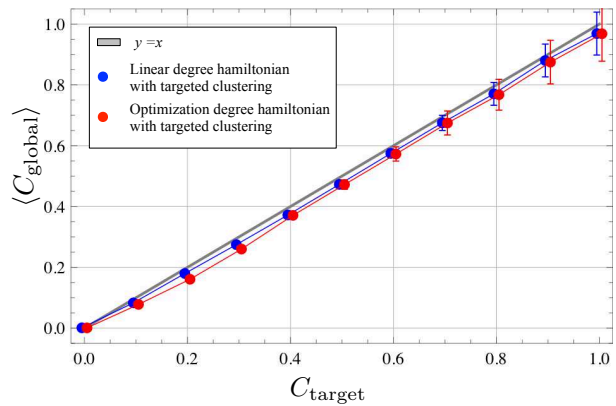}
\caption{The global clustering $\av{\Cg}$ of graph ensembles generated from the linear (blue) and the optimization (red) degree Hamiltonians perturbed with targeted clustering Hamiltonian $H_C=\sum_{i\in\cN}|t_i-\Ct s(k_i)|$. $\av{\Cg}\simeq\Ct$ is the result of node-level local clustering coefficients being $\simeq\Ct$, regardless of the degree distribution.}
\label{02_clustering}
\end{figure}

\begin{figure*}[t]
\includegraphics[width=180mm]{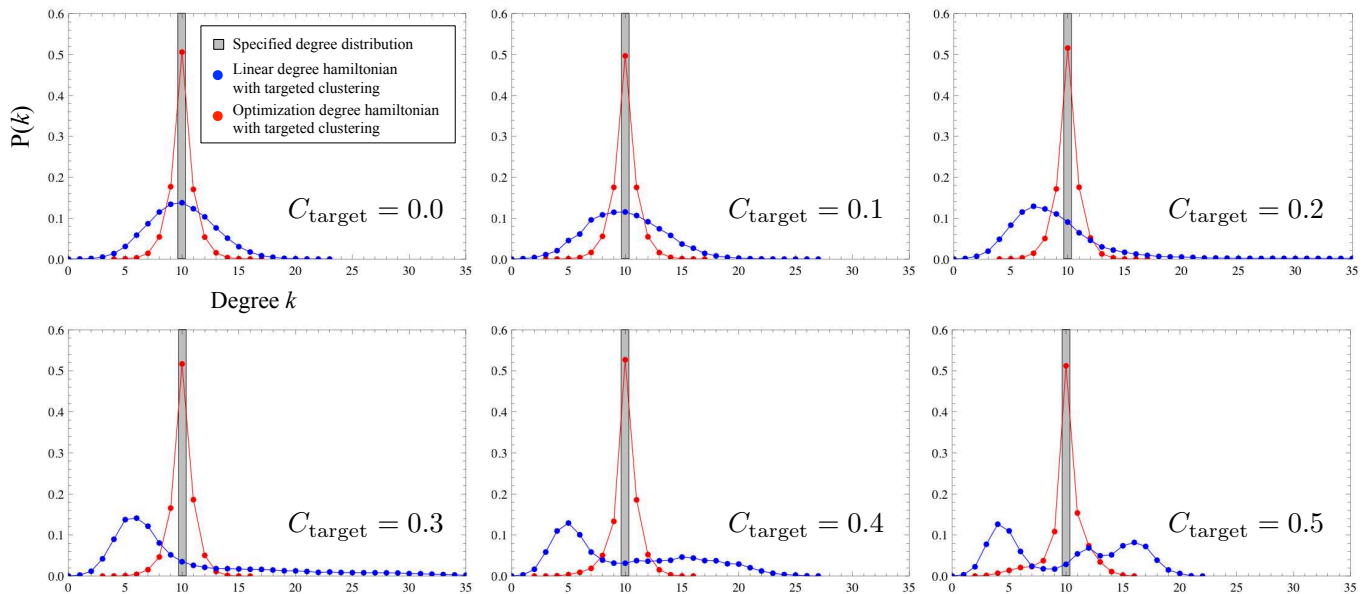}
\caption{The equilibrium degree distributions $P(k)$ generated from $H_1=\hld+\hc$ (blue) and $H_2=\hod+\hc$ (red) for various values of $\Ct$. The specified degree distribution $P(q)=\delta_{q,10}$ is in gray. When $\Ct=0$, both Hamiltonians generate their natural $P(k)$ -- a true Poissonian for $\hld$, and a sharp peak for $\hod$. As $\Ct$ is tuned higher, however, $P(k)$ peaks at a smaller $k$ for $\hld+\hc$ and even exhibits multiple peaks when $\Ct$ is too large, while it stays virtually unchanged for $\hod+\hc$ up to $\Ct\simeq0.5$, an unusually high value in real networks.}
\label{03_simdegree}
\end{figure*}

A Monte Carlo simulation was performed for a network of size $n=500$ and $\av{k}=10$. For simplicity, we again set $\beta_d=\beta_c=1$, $P(q)=\delta_{q,10}$ (a regular graph), and $\gi=\Ct$, a universal value for all $i$, varied between $0$ and $1$. First, Fig.~\ref{02_clustering} shows the mean global clustering $\av{\Cg}$ from the simulation, which shows us that both $H_1$ and $H_2$ $\av{\Cg}\simeq\Ct$ generate networks with the specified clustering. This arises from the fact that $\av{C_i}\simeq\Ct$ on the individual level as well (not shown).  The difference between the $H_1$ and $H_2$, however, is most striking in the equilibrium $P(k)$, shown in Fig.~\ref{03_simdegree}. When $\Ct=0$, perturbation $H_C$ is insignificant since the expected clustering without it is $0$ anyway, and therefore $P(k)$ are simply as expected -- a true Poissonian for $H_1$, and a sharper peak at $q=10$ for $H_2$, similar to the ones we saw in Fig.~\ref{01_degree}.  When $\Ct\ne0$, on the other hand, the peak in $P(k)$ under $H_1$ gradually shifts towards a smaller $k$ while high-degree nodes are created in order to compensate for the number of edges $M$ which is a constant. As $\Ct$ is tuned higher it resembles the specified distribution less and less, and at $\Ct\simeq0.4$ we even observe multiple peaks (at $k=5$ and $15$ -- the values for which $|t-\Ct s(k)|=0$, meaning the peaks will shift for a different $\Ct$ and thus are not very meaningful). $P(k)$ under $H_2$, in contrast, is robust, without noticeable change up to $\Ct=0.5$, already an unusually high value for real-world networks, until it too shows similar (but milder) behavior at a higher value of $\Ct\sim 0.6$ and up~\footnote{We performed similar simulations for the four heterogeneous $P(q)$ shown in Fig.~\ref{01A_smoothdegree} and found similar results.}.

Let us now discuss the implications of the findings in Figs.~\ref{02_clustering}~and~\ref{03_simdegree} on the topology of networks generated from $H_1=\hld+\hc$ and $H_2=\hld+\hc$. First of all, Fig.~\ref{02_clustering} tells us that, unlike the Strauss clustering perturbation $\tau T$, $\hc$ was able to discourage an extreme condensation of triangles, resulting in $\av{\Cg}\simeq\Ct$ by way of $C_i\simeq\Ct$ for both $H_1$ and $H_2$. However it was not enough to completely overcome the cooperative tendency of triangles under $\hld$. The telltale sign of this is the creation of high-degree nodes Fig.~\ref{03_simdegree} which shows the creation of high-degree nodes  in $H_1$: now many triangles exist between the high-degree nodes, forming a core of densely interconnected high-degree nodes although $C_i\simeq\Ct$ as specified. On the other hand, under $H_2$ where $P(k)$ is sharply peaked at the specified degree $q=10$ such cores does not exist; with $k_i\simeq q$ and $C_i\simeq\Ct$ for all $i$ as specified, $H_2$ generates a network that truly has a uniform distribution of triangles, lacking any unspecified, accidental local structures.

\begin{figure*}[t]
\includegraphics[width=180mm]{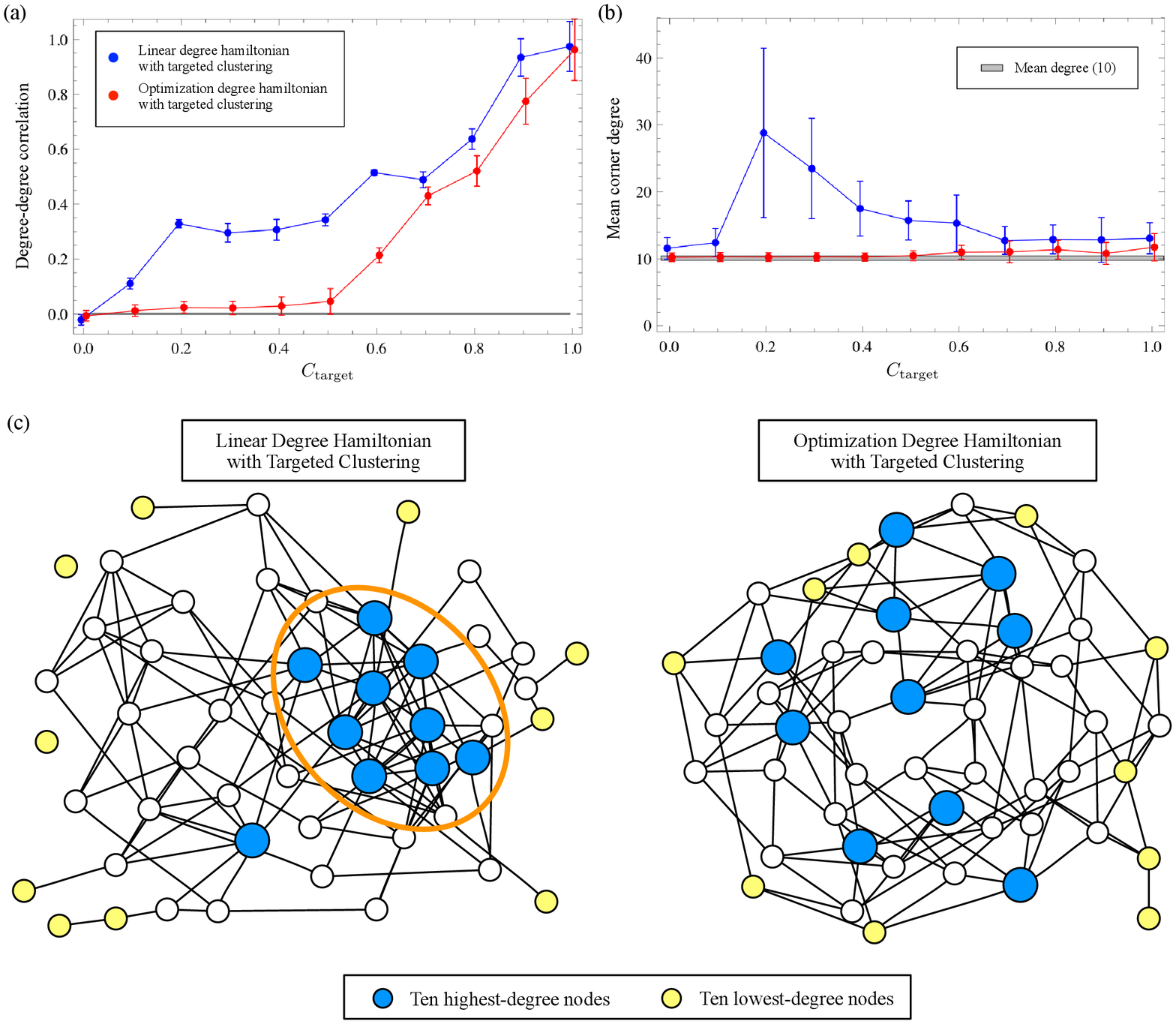}
\caption{(a) The degree-degree correlation $\rk$ under $H_1=\hld+\hc$ and $H_2=\hod+\hc$. $H_1$ generates positive degree correlation for any positive $\Ct$, while $H_2$ exhibits very little correlation up to $\Ct\simeq0.5$. (b) The mean corner degree of triangles contained in the networks in equilibrium. Under $H_1$ most triangles exist between high-degree nodes, indicating the persistence of the cooperative nature of triangles. (c) Equilibrium topologies of clustered networks under $H_1$ (left) and $H_2$ (right). $H_1$ generates a core of high-degree nodes that are densely connected and share a large number of triangles (enclosed in orange oval). $H_2$, in contrast, maintains the specified degree distribution $P(q)=\delta_{q,10}$ while the triangles are distributed uniformly, features expected of a maximally random configuration given the the degree and local clustering constraints.}
\label{04_degcortopo}
\end{figure*}

We check our claim via the following two quantities: the degree-degree correlation $\rk$ (the Pearson correlation between the degrees of adjacent nodes) and the \defn{mean corner degree} of the triangles in the network, shown in Figs~\ref{04_degcortopo}~(a)~and~(b). First, the plot of $\av{\rk}$ in Fig.~\ref{04_degcortopo}~(a) indicates  that adjacent degrees in the network are highly correlated under $H_1$, so that high-degree nodes are indeed connected with other high-degree nodes and vice versa, while $H_2$ shows no such effect. This leads naturally to what we see in Fig.~\ref{04_degcortopo}~(b): under $H_1$, the mean corner degree is significantly higher than $\av{k}=q$, unlike $H_2$ where it is practically equal to $q$. These observations are presented visually in Fig.~\ref{04_degcortopo}~(c) (an actual snapshot of an equilibrium configuration of a network with $n=50$, $P(q)=\delta_{q,5}$, and $\Ct=0.4$. $\av{\Cg}$ are $0.35\pm0.02$ and $0.31\pm0.02$, respectively). As expected, for $H_1$ (left) we clearly see that the ten highest-degree nodes (blue, average degree $9.7$) forms a densely interconnected core (encircled in orange), with ten lowest-degree nodes (yellow, average degree $1.0$) pushed to the periphery with low triangle participation rate. For $H_2$ (right), no significant difference between highest- and lowest-degree nodes exists, and the triangles are distributed uniformly, expected of a maximally random configuration given the degree and local clustering constraints.

\section{Discussion and Future Directions}
Here we have studied two forms of graph Hamiltonian in exponential random graph theory that take node degrees and local clustering as specified input. The tendency of triangles to coalesce in the Strauss model was shown to persist when the linear clustering perturbation was replaced by an optimized clustering form, albeit in a milder fashion, rendering the composite Hamiltonian unable to generate the specified degree distribution~\footnote{The reverse case of $\hod+\tau T$ was also numerically studied with varying $\tau$. As $\tau$ is tuned more negative (thus favoring triangles) here also occurs a sudden onset of the emergence of a densely connected cluster of high-degree nodes, signifying a condensation of triangles similar to $\hld+\tau T$. This demonstrates that to create finite clustering degree and local clustering optimization are necessary.}. The optimization degree Hamiltonian, on the other hand, was able to satisfy both, exhibiting significant robustness under the same perturbation.

That the optimization Hamiltonian form was able to reproduce both the targeted degree and clustering presents an appealing possibility from the viewpoint of network modeling via exponential random graph theory: given a set of network variables $\Phi=\set{\phi_v|v=1,\cdots,l}$, it may act as a practical computational method to generate a null model of network data with actual values of the variables $\set{\tilde{\phi}_v|v=1,\ldots,l}$ using the Hamiltonian~\cite{Foster:2010}
\begin{align}
H(G) = \sum_{\phi\in\Phi}\beta_v|\phi_v(G)-\tilde{\phi}_v|,
\label{genoptham}
\end{align}
thereby enabling the modeler to assess quickly the sufficiency of the particular set of variables in characterizing the network. An interesting recent application of a related framework was provided by Foster~\etal~\cite{Foster:2011}: specifically, they generated networks with specified global clustering coefficient $\Cg$ or degree-degree correlation $r$ using the optimization hamiltonian and measured their effect on each other and the modular structure of the network (although they kept the the degree sequence fixed as the network data). In doing so, they demonstrated the utility of the Hamiltonian of the form Eq.~\eref{genoptham} in creating network ensembles with desired characteristics. Naturally, more study must be made on the properties of Eq.~\eref{genoptham} in relation to various network variables --- global as well as local --- in order to establish its general utility, and also the more recent. In light of the fact that new, complex measures of network properties are frequently devised and introduced, we hope the formalism proves to be a useful tool for network scientists.

\begin{acknowledgments}
The author would like to thank Doochul Kim for helpful comments. This work was supported by Kyung Hee University Grant KHU-20100116 and the Korea Research Foundation Grant KRF-20100004910.
\end{acknowledgments}

\end{document}